\documentclass[prc,aps,twocolumn,showpacs,amssymb,superscriptaddress]{revtex4}

\usepackage{amsmath}
\usepackage{graphicx}
\usepackage{dcolumn}
\usepackage{float}
\begin{document}

\title{The nuclear matter equation of state with consistent two- and three-body
  perturbative chiral interactions}

\author{L. Coraggio}
\affiliation{Istituto Nazionale di Fisica Nucleare, \\
Complesso Universitario di Monte  S. Angelo, Via Cintia - I-80126 Napoli,
Italy}
\author{J. W. Holt}
\affiliation{Physics Department, University of Washington,
Seattle, WA 98195, USA}
\author{N. Itaco}
\affiliation{Istituto Nazionale di Fisica Nucleare, \\
Complesso Universitario di Monte  S. Angelo, Via Cintia - I-80126 Napoli,
Italy}
\affiliation{Dipartimento di Fisica, Universit\`a
di Napoli Federico II, \\
Complesso Universitario di Monte  S. Angelo, Via Cintia - I-80126 Napoli,
Italy}
\author{R. Machleidt}
\affiliation{Department of Physics, University of Idaho, Moscow, ID 83844, USA}
\author{L. E. Marcucci}
\affiliation{Dipartimento di Fisica ``Enrico Fermi'', Universit\`a
di Pisa, 
Largo Bruno Pontecorvo 3 - I-56127 Pisa, Italy}
\affiliation{Istituto Nazionale di Fisica Nucleare, Sezione di Pisa,\\
Largo Bruno Pontecorvo 3 - I-56127 Pisa, Italy}
\author{F. Sammarruca}
\affiliation{Department of Physics, University of Idaho, Moscow, ID 83844, USA}

\date{\today}

\begin{abstract}
We compute the energy per particle of infinite symmetric nuclear
matter from chiral N$^3$LO (next-to-next-to-next-to-leading order)
two-body potentials plus N$^2$LO three-body forces. 
The low-energy constants of the chiral three-nucleon force that cannot
be constrained by two-body observables are fitted to reproduce the
triton binding energy and the $^3$H-$^3$He Gamow-Teller transition
matrix element. 
In this way, the saturation properties of nuclear matter are
reproduced in a parameter-free approach.
The equation of state is computed up to third order in many-body
perturbation theory, with special emphasis on the role of the
third-order particle-hole diagram. 
The dependence of these results on the cutoff scale and regulator
function is studied. 
We find that the inclusion of three-nucleon forces consistent with the
applied two-nucleon interaction leads to a reduced dependence on the
choice of the regulator only for lower values of the cutoff.
\end{abstract}


\maketitle

\section{Introduction}
\label{intro}

High-precision nuclear potentials based on chiral perturbation theory (ChPT) 
\cite{EM03,EGM05,ME11} are nowadays widely employed to link the
fundamental theory of strong interactions, quantum chromodynamics
(QCD), to nuclear many-body phenomena.
An important feature of ChPT is that nuclear two-body forces, many-body 
forces, and currents \cite{Wei92,Kol94,ME11} are generated on an equal footing. 
Consistency then requires that certain low-energy constants (LECs)
appearing in the two-nucleon-force (2NF) --- and fitted to two-nucleon
data --- appear also in three-nucleon forces (3NF), four-nucleon
forces (4NF), and electroweak currents.

Since ChPT is a low-momentum expansion valid only for momenta $ Q <
\Lambda_\chi \simeq 1$ GeV, where $\Lambda_\chi$ denotes the chiral
symmetry breaking scale, nucleon-nucleon ($NN$) potentials derived
from ChPT are typically multiplied by a (non-local) regulator function
\begin{equation}
f(p',p) = \exp [-(p'/\Lambda)^{2n} - (p/\Lambda)^{2n}] \,,
\label{eq_reg}
\end{equation} 
where $\Lambda \simeq 0.5$ GeV is a typical choice for the cutoff scale.
In the effective field theory (EFT) framework, the calculated physical
observables ideally will be independent of both the regulator function
and the associated cutoff scale $\Lambda$. In the case of nuclear
interactions this is rarely the case, and varying the regulator is
often used as a tool to estimate the uncertainty in the theoretical
calculations. 
In the two-nucleon problem, the dependence of the solutions of the
Lippmann-Schwinger equation on the regulator function and its cutoff
scale is minimized by a renormalization procedure in which the LECs
associated with $NN$ and $\pi N$ vertices are readjusted to
two-nucleon phase shifts and deuteron properties. 
Even though potentials with different regulator functions yield
similar phase shifts, they will in general give different predictions
when employed in many-body calculations, due to their different
off-shell behavior.
One is then faced with a larger cutoff dependence in many-body systems
\cite{Cor12}, which should be reduced by a consistent adjustment of
the LECs appearing in nuclear many-body forces.

In a recent paper \cite{Coraggio13}, we have studied the regulator
dependence of the cold neutron matter equation of state (EOS)
employing chiral two- and three-nucleon potentials within many-body
perturbation theory. 
Previous studies of infinite symmetric nuclear matter and pure neutron
matter
\cite{Sammarruca12,Tews13,HKW13,Krueger13,Gezerlis13,Carbone13,Baardsen13,Kohno13,Hagen14}
have focused on the importance of nuclear many-body forces and have
explored the perturbative and non-perturbative features of chiral
nuclear potentials.
In Ref.\ \cite{Coraggio13} we observed that in neutron matter
calculations the use of consistent 3NFs plays a crucial role in the
restoration of regulator independence.
The calculation of the ground state energy of infinite neutron matter
with chiral 3NFs up to N$^2$LO depends only on LECs that have been
fixed in the two-nucleon system \cite{HS10}. In the case of symmetric
nuclear matter, also the one-pion exchange three-nucleon force
$V_{3N}^{1\pi}$ and the contact three-nucleon force $V_{3N}^{\rm
  cont}$ at N$^2$LO contribute. 
Therefore, the associated LECs $c_D$ and $c_E$, which are not
constrained by two-body observables, must also be refitted for
different regulator functions. 
These 3NF LECs should be adjusted to $A = 3$ observables only, and a
possible choice \cite{Gardestig06,Gazit08,Marcucci12} is to reproduce
the $^3$H and $^3$He binding energies together with the triton
half-life (specifically the Gamow-Teller matrix element).

In the present work, we continue the investigation started in Ref.\
\cite{Coraggio13} and study the dependence of the EOS of symmetric
nuclear matter on the choice of regulator function in the chiral
nuclear potentials, employing two- and three-nucleon forces with
consistent LECs. 
The ability to obtain realistic nuclear matter predictions with
(consistent) two- and three-body interactions constrained by the
properties of the two- and the three-nucleon systems and no additional
adjustments is a focal point of this paper.
Historically, this has proven to be a non-trivial task. As in Ref.\
\cite{Coraggio13}, we employ three different chiral potentials with
cutoff scales $\Lambda = 414$ MeV \cite{Cor07}, 450 MeV, and 500 MeV
\cite{EM03,ME11}. 
The LECs $c_D$ and $c_E$ of the N$^2$LO chiral three-nucleon force are
fitted, for each value of $\Lambda$, to the binding energies of $A=3$
nuclei and the $^3$H-$^3$He Gamow-Teller matrix element.
Note that the $\Lambda=500$ MeV two- and three-nucleon chiral
potentials have been used to study $A=3$ and 4 elastic scattering
\cite{Viv13}, the $A\leq 3$ nuclei electromagnetic structure
\cite{Pia13} and low energy reactions of astrophysical interest
\cite{Mar13}, finding good agreement with the experimental data when
available.

We compute the energy per particle of symmetric nuclear matter up to
third order in many-body perturbation theory.
Previous calculations \cite{Hebeler11,Carbone13,Baardsen13,Kohno13}
beyond second order have focused on the inclusion of particle-particle
($pp$) and hole-hole ($hh$) ladder diagrams, whereas in the present
work we compute, in addition to the third-order $pp$ and $hh$
diagrams, also the third-order particle-hole ($ph$) diagram (without
simplifying approximations), which has not been considered previously
but is necessary for a consistent third-order calculation.
The effects of the N$^2$LO 3NF are included via a density-dependent
two-body potential ${\overline V}_{3N}$ that is added to the chiral
N$^3$LO potential $V_{2N}$, and which is obtained by summing one
nucleon over the non-interacting filled Fermi sea of nucleons
\cite{HKW09,HS10,HKW10}.

The paper is organized as follows. 
In Sec.\ \ref{c23np}, we briefly describe the features of the
different chiral potentials employed and provide details about the
renormalization procedure we have followed to choose the LECs of the
3NF terms $V_{3N}^{1\pi}$ and $V_{3N}^{\rm cont}$.
In Sec.\ \ref{nmc}, we outline the perturbative calculation of the
energy per particle in symmetric nuclear matter that takes into
account 3NF effects.
Our results and conclusions are presented in Secs.\ \ref{res} and \ref{conc},
respectively.

\section{Scale dependence of chiral two- and three-nucleon potentials}
\label{c23np}
During the past two decades, chiral EFT has emerged as a powerful tool
for describing hadronic interactions at low energy scales in a
systematic and model-independent way (see Refs.\ \cite{ME11,EHM09} for
recent reviews). 
The separation of scales required to construct a useful EFT arises
naturally in nuclear interactions from the pseudo-Goldstone boson
nature of pions, which is associated with the spontaneous breaking of
chiral symmetry and is responsible for the large difference between
the light pion mass ($m_\pi \simeq$ 135 MeV) and the masses of the
next lowest states in the meson spectrum, the $\rho(770)$ and
$\omega(782)$.

\begin{figure}[t]
\begin{center}
\includegraphics[scale=0.45,angle=0]{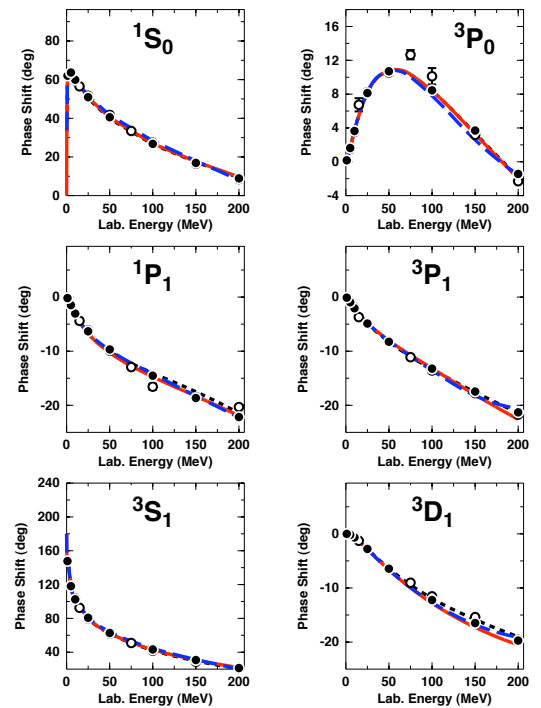}
\vspace*{-2cm}
\caption{(Color online) Neutron-proton phase shifts as predicted by
  chiral N$^3$LO potentials with different cutoff scale
  $\Lambda$. Solid (red) curve, $\Lambda=414$ MeV; dashed (blue)
  curve, $\Lambda=450$ MeV; and dotted (black) curve, $\Lambda=500$
  MeV. Partial waves with total angular momentum $J\leq 1$ are
  displayed. The solid dots and open circles are the results from the
  Nijmegen multi-energy $np$ phase shift analysis~\protect\cite{Sto93}
  and the VPI/GWU single-energy $np$ analysis
  SM99~\protect\cite{SM99}, respectively.}
\label{ph1}
\end{center}
\end{figure}

\begin{figure}[h]
\begin{center}
\includegraphics[scale=0.45,angle=0]{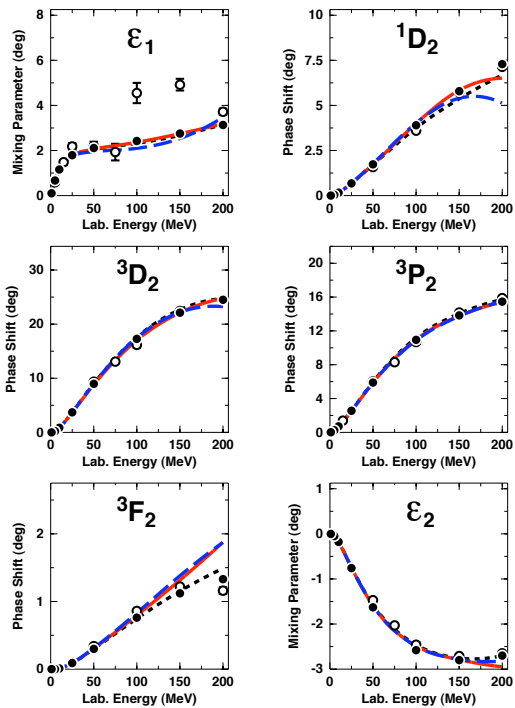}
\vspace*{-2cm}
\caption{(Color online) Same as Fig.~\ref{ph1}, but for $J=2$ phase
  shifts and $J\leq 2$ mixing parameters.}
\label{ph2}
\end{center}
\end{figure}

\begin{figure}[h]
\begin{center}
\includegraphics[scale=0.45,angle=0]{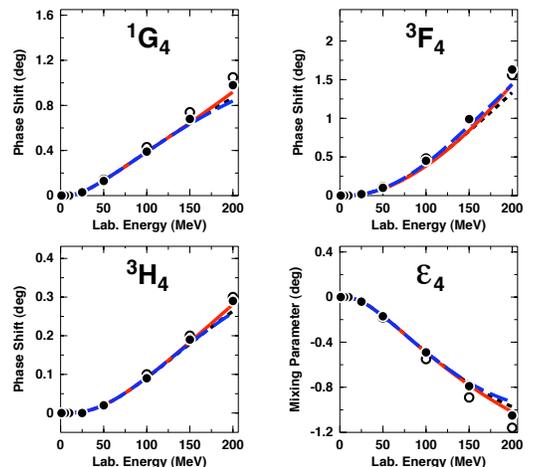}
\vspace*{-6cm}
\caption{(Color online) Same as Fig.~\ref{ph1}, but for some
  representative peripheral partial waves.}
\label{ph3}
\end{center}
\end{figure}

In normal nuclear many-body systems, the nuclear momenta are on the
order of the pion mass, and therefore the systematic construction of
chiral nuclear potentials is based on an expansion in powers of this
soft scale ($Q \sim m_\pi$) over the hard scale set by the typical
hadron masses $\Lambda_\chi \sim m_\rho \sim 1$~GeV, also known as the
chiral-symmetry breaking scale. 
For this EFT to rise above the level of phenomenology, it must have a
firm link with QCD.
The link is established by having the EFT observe all relevant
symmetries of the underlying theory, in particular, the broken chiral
symmetry of low-energy QCD~\cite{Wei79}.
The past 15 years have seen great progress in applying ChPT to nuclear
forces. 
As a result, $NN$ potentials of high precision have been constructed,
which are based upon ChPT carried to N$^3$LO.

Since ChPT is a low-momentum expansion, valid only for momenta $Q <
\Lambda_\chi$, the potentials are multiplied with a regulator
function, like, e.~g., the one of Gaussian shape given in
Eq.~(\ref{eq_reg}).
In this investigation, we consider three N$^3$LO $NN$ potentials which
differ by the cutoff scale $\Lambda$ and the regulator function: (i)
$\Lambda = 414$ MeV using the regulator function Eq.~(\ref{eq_reg})
with $n=10$, i.e., a smooth, but rather steep cutoff function is
applied.
This potential is very similar to the one with a sharp cutoff at 414
MeV published in Ref.~\cite{Cor07}; however, a smooth version of the
steep cutoff is more convenient in calculations of the three-body
system. 
(ii) $\Lambda = 450$ MeV, using the regulator function
Eq.~(\ref{eq_reg}) with $n=3$, which has been constructed for our
study of Ref.~\cite{Coraggio13} and the present investigation; (iii)
$\Lambda = 500$ MeV, using the regulator function Eq.~(\ref{eq_reg})
with $n=2$ for the 2$\pi$ exchange contributions~\cite{EM03}.
All three potentials use the same (comprehensive) analytic expressions
which can be found in Ref.~\cite{ME11}.
Note that the Gaussian regulator function of Eq.~(\ref{eq_reg})
suppresses the potential also for $Q<\Lambda$, particularly for small
$n$, which is the reason why we use $n=10$ for the case of the lowest
cutoff of 414 MeV.
Cutoff-independence is an important aspect of an EFT.
In lower partial waves, the cutoff dependence of the $NN$ phase shifts
is counterbalanced by an appropriate adjustment of the contact terms
which, at N$^3$LO, contribute in $S$, $P$, and $D$ waves.
The extent to which cutoff independence can be achieved in lower
partial waves is demonstrated in Figs.~\ref{ph1} and \ref{ph2}.
In $F$ and higher partial waves (where there are no $NN$ contact
terms) the LECs of the dimension-two $\pi N$ Lagrangian can be used to
obtain cutoff independence of the phase shift predictions, as shown in
Fig.~\ref{ph3}.

\begin{figure}[b]
\begin{center}
\includegraphics[scale=0.4,angle=0]{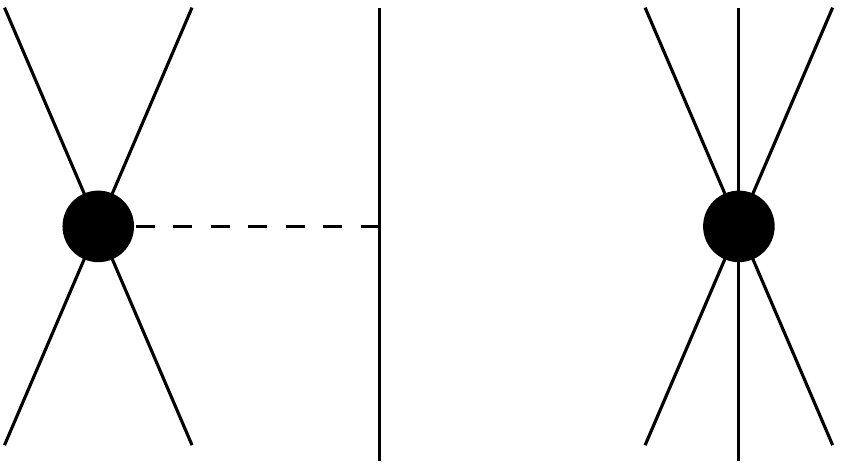}
\caption{The N$^2$LO three-nucleon force contact interactions:
  $V_{3N}^{1\pi}$ on the left and $V_{3N}^{\rm cont}$ on the right
  (see Eqs.~(\ref{eq_3nf_nnloc}) and (\ref{eq_3nf_nnlod}), respectively).}
\label{fig:contact}
\end{center}
\end{figure}

An important advantage of the EFT approach to nuclear forces is that
it creates two- and many-body forces on an equal footing.
The first non-vanishing 3NF occurs at N$^2$LO. 
At this order, there are three 3NF topologies: the two-pion exchange
(2PE), one-pion exchange (1PE) plus a 2N-contact interaction, and a
pure 3N-contact interaction. 
These last two topologies are represented in Fig.~\ref{fig:contact}.

The 2PE 3N-potential is given by
\begin{equation}
V_{3N}^{2\pi} = 
\left( \frac{g_A}{2f_\pi} \right)^2
\frac12 
\sum_{i \neq j \neq k}
\frac{
( \vec \sigma_i \cdot \vec q_i ) 
( \vec \sigma_j \cdot \vec q_j ) }{
( q^2_i + m^2_\pi )
( q^2_j + m^2_\pi ) } \;
F^{ab}_{ijk} \;
\tau^a_i \tau^b_j
\label{eq_3nf_nnloa}
\end{equation}
with $\vec q_i \equiv \vec{p_i}' - \vec p_i$, where $\vec p_i$ and
$\vec{p_i}'$ are the initial and final momenta of nucleon $i$,
respectively, and
\begin{equation}
F^{ab}_{ijk} = \delta^{ab}
\left[ - \frac{4c_1 m^2_\pi}{f^2_\pi}
+ \frac{2c_3}{f^2_\pi} \; \vec q_i \cdot \vec q_j \right]
+ 
\frac{c_4}{f^2_\pi}  
\sum_{c} 
\epsilon^{abc} \;
\tau^c_k \; \vec \sigma_k \cdot [ \vec q_i \times \vec q_j] \; .
\label{eq_3nf_nnlob}
\end{equation}  
Note that the 2PE 3NF does not contain any new parameters, because the
LECs $c_1$, $c_3$, and $c_4$ appear already  in the 2PE 2NF.
The 1PE contribution is
\begin{equation}
V_{3N}^{1\pi} = 
-\frac{c_D}{f^2_\pi\Lambda_\chi} \; \frac{g_A}{8f^2_\pi} 
\sum_{i \neq j \neq k}
\frac{\vec \sigma_j \cdot \vec q_j}{
 q^2_j + m^2_\pi }
( \mbox{\boldmath $\tau$}_i \cdot \mbox{\boldmath $\tau$}_j ) 
( \vec \sigma_i \cdot \vec q_j ) 
\label{eq_3nf_nnloc}
\end{equation}
and the 3N contact potential reads
\begin{equation}
V_{3N}^{\rm cont} = 
\frac{c_E}{f^4_\pi\Lambda_\chi}
\; \frac12
\sum_{j \neq k} 
 \mbox{\boldmath $\tau$}_j \cdot \mbox{\boldmath $\tau$}_k  \; .
\label{eq_3nf_nnlod}
\end{equation}
In the above, we use $g_A=1.29$, $f_\pi=92.4$ MeV, $m_\pi=138.04$ MeV,
and $\Lambda_\chi=700$ MeV.

\begin{figure}[tb]
\begin{center}
\includegraphics[scale=0.3,angle=0]{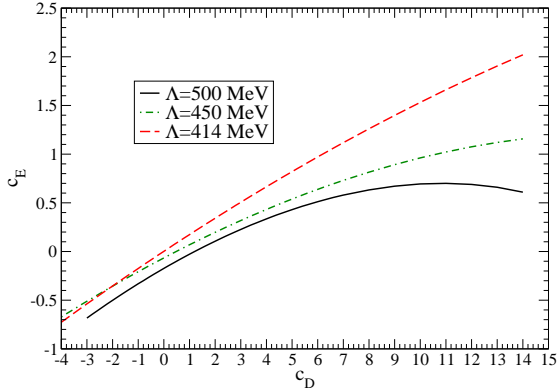}
\caption{(Color online) $c_D$-$c_E$ trajectories fitted to reproduce
  the experimental $^3$H and $^3$He binding energies. Solid (black)
  curve for $\Lambda=500$ MeV, dotted-dashed (green) curve for
  $\Lambda=450$ MeV, and dashed (red) curve for $\Lambda=414$ MeV.}
\label{fig:cdce}
\end{center}
\end{figure}
 
\begin{figure}[tb]
\begin{center}
\includegraphics[scale=0.3,angle=0]{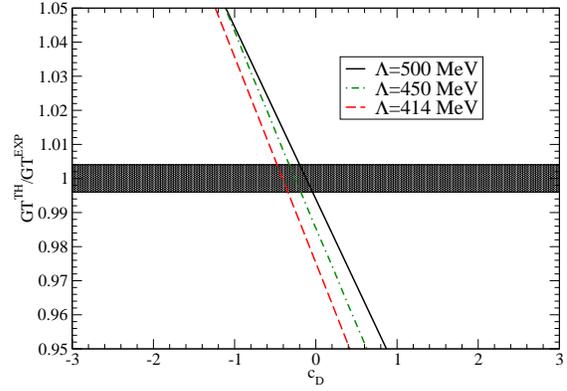}
\caption{(Color online) The ratio between the calculated GT value
  (GT$^{\rm TH}$) and the experimental one (GT$^{\rm EXP}$) as a
  function of the LEC $c_D$. Solid (black) curve for $\Lambda=500$
  MeV, dotted-dashed (green) curve for $\Lambda=450$ MeV, and dashed
  (red) curve for $\Lambda=414$ MeV. The shaded stripe represents the
  experimental uncertainty.}
\label{fig:gt}
\end{center}
\end{figure}

The last two 3NF terms involve the two new parameters $c_D$ and $c_E$,
which do not appear in the 2N problem.
There are many ways to constrain these two parameters.
The triton binding energy and the $nd$ doublet scattering length
$^2a_{nd}$ or the $^4$He binding energy can be used.
Given the known correlation between these observables, one may choose
instead an optimal over-all fit of the properties of light
nuclei~\cite{Nav07}. 
However, recently a new procedure has been used to fix $c_D$ and $c_E$
\cite{Gardestig06,Gazit08,Marcucci12}.
Due to the consistency of interactions and currents in chiral
EFT~\cite{Gardestig06,Gazit08}, the LEC $c_D$ that appears in
$V_{3N}^{1\pi}$ is also involved in the two-nucleon contact term in
the $NN$ axial current operator derived up to N$^2$LO. 
Therefore, $c_D$ can be constrained using the accurate experimental
value of one observable from weak processes involving two- or
few-nucleon systems.
Given the lack of accurate experimental values for weak observables in
the two-body sector, the choice has been to use the triton
$\beta$-decay half-life, in particular its Gamow-Teller (GT)
component. 
This observable has been used already in a variety of studies to
constrain the two-body axial current
operator~\cite{Mar00,Par03,Mar11,Marcucci12,Mar13}.
Therefore, we proceed here as in Ref.~\cite{Marcucci12}: 
(i) We calculate the $^3$H and $^3$He wave functions with the
hyperspherical harmonics method (see Ref.~\cite{Kie08} for a review),
using the chiral 2NF plus 3NF presented above for each cutoff
parameter $\Lambda$. 
The corresponding set of LECs $c_D$ and $c_E$ are determined by
fitting the $A=3$ experimental binding energies.
The resulting trajectories are shown in Fig.~\ref{fig:cdce}. 
(ii) For each set of $c_D$ and $c_E$, the $^3$H and $^3$He wave
functions are used to calculate the GT matrix element. 
Comparison with the experimental value leads to a range of values for
$c_D$ for each cutoff parameter $\Lambda$, as shown in
Fig.~\ref{fig:gt}. 
Then, from Fig.~\ref{fig:cdce}, the corresponding range for $c_E$ is
determined. 
The values for $c_D$ and $c_E$ used in the present calculation are
listed in Table~\ref{tab1} for each $\Lambda$.
Note that the values of $c_D$ and $c_E$ for $\Lambda=500$ MeV are
slightly different from those used in previous studies
\cite{Marcucci12,Viv13,Mar13}, but the GT matrix element is still
reproduced within less than 1 \%. 

\begin{table*}
\caption{For the various chiral N$^3$LO $NN$ potentials used in the
  present investigation, we list the cutoff $\Lambda$, the
  type of regulator, the exponent $n$ used in the regulator function
  [see Eq.~(\ref{eq_reg})], the LECs  of the dimension-two $\pi N$
  Lagrangian, $c_i$ (in units of GeV$^{-1}$), and the LECs $c_D$ and
  $c_E$ entering the three-nucleon potential.}
\label{tab1}
\smallskip
\begin{tabular*}{\textwidth}{@{\extracolsep{\fill}}cccc}
\hline
\hline
\noalign{\smallskip}
               & \multicolumn{3}{c}{Cutoff parameter $\Lambda$ (MeV)} \\
               \cline{2-4}
                 &      414                     & 450           &    500            \\
\hline
\noalign{\smallskip}
Regulator type & Gaussian & Gaussian & Gaussian \\
$n$ & 10 & 3 & 2 \\
$c_1$ & --0.81 & --0.81 & --0.81 \\
$c_2$ &   3.28 &   3.28 &   2.80 \\
$c_3$ & --3.00 & --3.40 & --3.20 \\
$c_4$ &   3.40 &   3.40 &   5.40 \\
$c_D$ & --0.40 & --0.24 &   0.0  \\
$c_E$ & --0.07 & --0.11 & --0.18 \\
\hline
\hline
\noalign{\smallskip}
\end{tabular*}
\end{table*}
 
At this point, a final remark is in order: the present fitting procedure employs
N$^3$LO $NN$ interaction together with a 3NF derived at N$^2$LO, i.e.,
one chiral order lower. 
Furthermore, also the weak current operator is derived at N$^2$LO. 
To avoid this mismatch, it would be necessary to
use N$^3$LO two- and three-nucleon interactions and currents. 
However, a derivation at N$^3$LO of the electroweak current would
require the calculation of loop corrections, a task successfully
performed for the electromagnetic~\cite{Pia13}, but not yet for the
axial operators.
Furthermore, the 3NF at N$^3$LO has been derived only recently ~\cite{Ber08,Ber11} 
and, due to its complexity, only preliminary
studies have been performed
so far~\cite{Hagen14,Ski13,Wit13}.
 
\section{Nuclear matter calculations}
\label{nmc}
We calculate the ground state energy (g.s.e.) per particle of infinite
symmetric nuclear matter within the framework of many-body
perturbation theory.
More precisely, the g.s.e.\ is expressed as a sum of Goldstone
diagrams up to third order.

As mentioned in Sec.~\ref{intro}, the effects of the 3NF are
taken into account via a density-dependent two-body potential
${\overline V}_{3N}$, that is added to the chiral N$^3$LO potential $V_{2N}$.
This potential is obtained by integrating one nucleon up to the Fermi
momentum $k_F$, thus leading to a density-dependent two-nucleon
interaction ${\overline V}_{3N}(k_F)$.
At this time, analytic expressions for ${\overline V}_{3N}$
\cite{HKW09,HKW10} has been derived only for the N$^2$LO 3NF, which is
the one we take into account in this work.
We recall that to take care of the correct combinatorial factors of
the normal-ordering at the two-body level of the 3NF, the matrix
elements of ${\overline V}_{3N}(k_F)$ have to be multiplied by a
factor 1/3 in the first-order Hartree-Fock (HF) diagram, and by a
factor 1/2  in the calculation of the HF single-particle energies
\cite{HS10,Coraggio13}.

\begin{figure}[H]
\begin{center}
\includegraphics[scale=0.8,angle=-90]{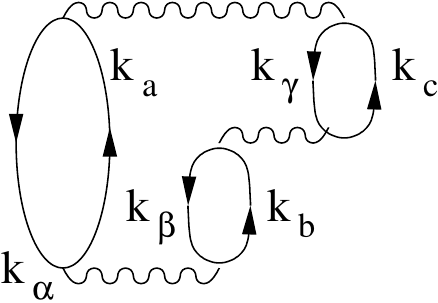}
\caption{Third-order ring diagram of the Goldstone expansion that we
  have included in our calculations with $V_{2N}$ and ${\overline
    V}_{3N}$ vertices. Latin-letter subscripts denote particle states,
  Greek-letter subscripts correspond to hole states.}
\label{figring}
\end{center}
\end{figure}

We point out that in the present calculations we have summed the
perturbation expansion up to third-order in $V_{2N}+{\overline
  V}_{3N}$, in particular including the third-order particle-hole
($ph$) diagram (see Fig. \ref{figring}), which is also known as the
third-order ring diagram \cite{Bethe65}.
This diagram has been taken into account neither in our previous paper
\cite{Coraggio13}, nor in other recent nuclear matter calculations
which have employed chiral potentials within a perturbative approach 
\cite{Sammarruca12,Tews13,Krueger13,Kohno13}.
The analytic expressions of first-, second-, and third-order
particle-particle ($pp$) and hole-hole ($hh$) contributions, together
with the one of single-particle HF potential, have been already
reported in Ref. \cite{Coraggio13}.
The implicit expression of the third-order $ph$ diagram can be found
in Ref. \cite{MacKenzie69}, where also the explicit expressions for a
potential without tensor and spin-orbit forces are reported.

The contributions of each diagram to the perturbation expansion obtained 
with the three chiral potentials for $k_F=1.3$ fm$^{-1}$ without and with 
3NF effects are given in Tables \ref{tab2} and \ref{tab3}, respectively.
It is clear that the magnitude of the third-order $ph$ diagram is
large, bringing a relevant contribution to the third-order energy.

This is in line with the results shown in Ref. \cite{Hagen14}, where
the neutron and nuclear EOS have been calculated within the
coupled-cluster approach employing the chiral NNLO$_{\rm opt}$
potential \cite{Ekstrom13}.
As a matter of fact, in \cite{Hagen14} the inclusion of perturbative
triples corrections in the coupled-cluster equations leads to
corrections for the binding energy of about 1 MeV per nucleon, when
including the 3NF in the normal-ordered two-body approximation.
Furthermore, it is insightful to note that in Ref.\cite{Hagen14}
a significant contribution was found when going beyond the 3NF
normal-ordered two-body approximation. 
With that in mind, we estimate the uncertainty of our perturbative
result to be approximately 2 MeV. 

In order to study the convergence properties of the perturbative
expansion, it is useful to consider the $[2|1]$ Pad\'e approximant \cite{BG70}

\begin{equation}
E_{[ 2|1 ]}=\mathcal{E}_0 + \mathcal{E}_1 +
\frac{\mathcal{E}_2}{1-\mathcal{E}_{3}/\mathcal{E}_2}~~,
\end{equation}

\noindent
$\mathcal{E}_i$ being the $i$th order energy contribution in the
perturbative expansion of the g.s.e..
The Pad\'e approximant is an estimate of the value to which the
perturbative series may converge.
Thus, in the following section we will perform a comparison between
the third-order results and those obtained by means of the $[2|1]$
Pad\'e approximant, to obtain an indication of the size of the
higher-order perturbative terms.

\section{Results}
\label{res}
In this section, we report the results of the calculation of the EOS
of infinite symmetric nuclear matter in the framework of many-body
perturbation theory. 
Since we include all contributions up to third order in the
interaction, we are in a good position to study the convergence
properties of the perturbative expansion.

\begin{figure}[tb]
\begin{center}
\includegraphics[scale=0.35,angle=0]{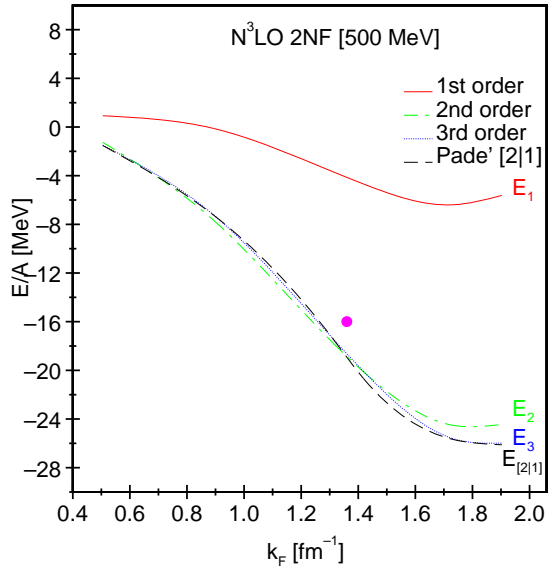}
\caption{(Color online) Nuclear matter energy per particle obtained
  from the N$^3$LO 2NF with cutoff $\Lambda=500$ MeV. The first,
  second, and third order in the perturbative expansion and the Pad\'e
  approximant $[2|1]$ are shown as a function of the Fermi momentum $k_F$.}
\label{conv1}
\end{center}
\end{figure}

\begin{figure}[tb]
\begin{center}
\includegraphics[scale=0.35,angle=0]{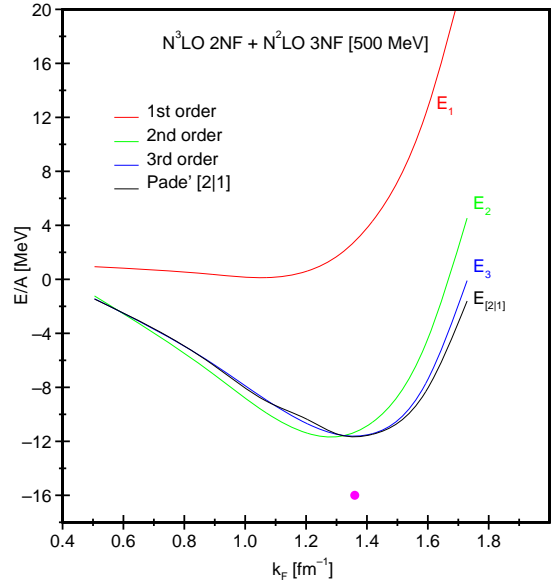}
\caption{(Color online) Same as in Fig. \ref{conv1}, but including the
  contribution of the N$^2$LO 3NF.}
\label{conv2}
\end{center}
\end{figure}

We find that among the three chiral potentials under consideration,
the least satisfactory perturbative behavior belongs to the chiral
N$^3$LO $NN$ potential with $\Lambda=500$ MeV, whether the
corresponding N$^2$LO 3NF is included or not.
This feature was already observed in our study of pure neutron matter
\cite{Coraggio13} and is apparent in Figs.~\ref{conv1}
and~\ref{conv2}.
In Fig.~\ref{conv1} we show the EOS as a function of the Fermi
momentum $k_F$, calculated at various orders in the perturbative
expansion applying the chiral N$^3$LO $NN$ potential with
$\Lambda=500$ MeV.
By inspection of the figure, it can be seen that the energy per
nucleon calculated at second order, $E_2$, does not differ much from
the one computed at third order, $E_3$, for the whole range of Fermi
momenta considered.
The perturbative character is also indicated by the fact that the
curve corresponding to $E_3$ is almost indistinguishable from the
$[2|1]$ Pad\'e approximant one.

\begin{figure}[tb]
\begin{center}
\includegraphics[scale=0.35,angle=0]{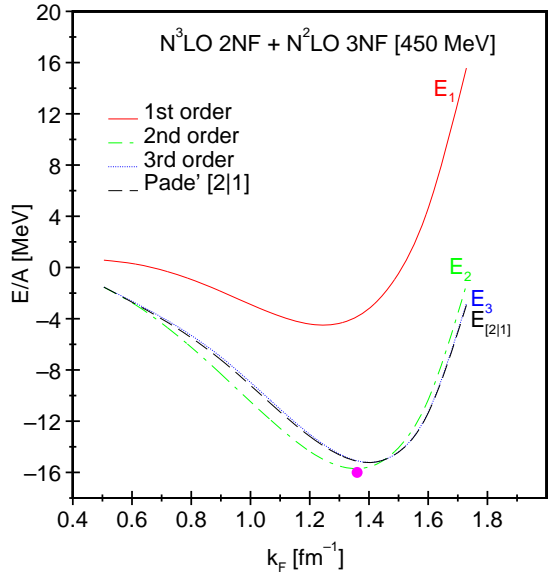}
\caption{(Color online) Same as Fig.~\ref{conv2}, but for
  $\Lambda=450$ MeV.}
\label{conv_450}
\end{center}
\end{figure}

Different considerations about the perturbative expansion have to be
drawn when including the effects of 3NF.
As a matter of fact, from inspection of Fig.~\ref{conv2}, it can be
seen that now the curve corresponding to $E_3$ deviates from the one
given by the $[2|1]$ Pad\'e approximant for $k_F$ larger than 1.6
fm$^{-1}$, indicating a worsening of the perturbative behavior.
On the other hand, using the other chiral potentials with lower
cutoffs, the perturbative behavior is satisfactory at least up to
$k_F=1.8$ fm$^{-1}$, as shown in Fig.~\ref{conv_450} for $\Lambda=450$
MeV.

\begin{figure}[tb]
\begin{center}
\includegraphics[scale=0.35,angle=0]{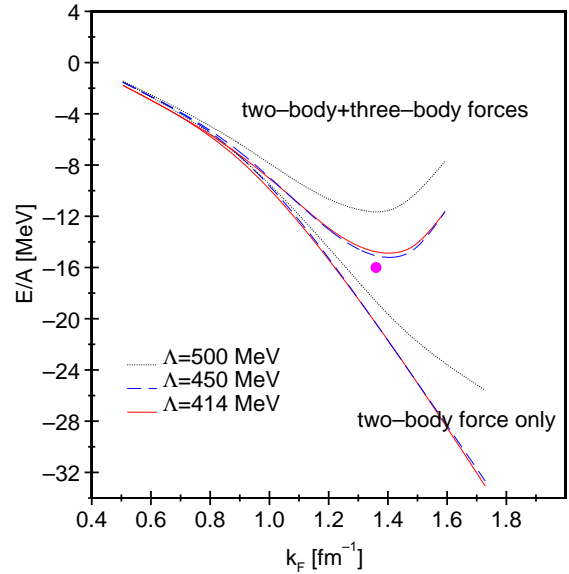}
\caption{(Color online) Results obtained for the g.s.e. per particle
  of infinite nuclear matter at third-order in perturbation theory
  for three sets of chiral interactions which differ by the cutoff
  $\Lambda$.}
\label{comparison}
\end{center}
\end{figure}

In Fig.~\ref{comparison} we display our predicted EOS obtained with
chiral potentials that apply different regulator functions.
We have added to each 2NF a 3NF whose LECs $c_i$, cutoff parameters,
and regulator function are exactly the same as in the corresponding
N$^3$LO $NN$ potential, see Table~\ref{tab1}, while the $c_D$ and
$c_E$ LECs have been chosen such as to reproduce the observed $A=3$
binding energies and triton Gamow-Teller matrix element (see Sec.~II).
Our results have been obtained at third-order in the perturbative
expansion, with and without taking into account 3NF effects.

The EOS calculated with 2NFs only and cutoffs $\Lambda=414,~450$ MeV
are very close to each other, while the one corresponding to
$\Lambda=500$ MeV is very different from the others.
None of them show saturation, at least up to $k_F=1.9$ fm$^{-1}$.
The differences between the predictions obtained with the two lower
cutoffs on the one hand, and with the larger cutoff on the other, are
not removed when including three-body effects.
As a matter of fact, while the $\Lambda=414$ and 450 MeV EOS are
nearly identical and show realistic saturation properties, the
$\Lambda=500$ MeV EOS is considerably more repulsive.
This is quite different from what we observed in pure neutron matter,
where the inclusion of 3NF effects resulted in a (net) strong
regulator-dependence reduction, with the predictions from the three
potentials approaching one another.

\begin{table*}
\caption{Contributions of each diagram to the perturbation expansion
  (in MeV) obtained with the three chiral potentials for $k_F=1.3$
  fm$^{-1}$ taking into account only 2NFs.}
\label{tab2}
\smallskip
\begin{tabular*}{\textwidth}{@{\extracolsep{\fill}}cccc}
\hline
\hline
\noalign{\smallskip}
               & \multicolumn{3}{c}{Cutoff parameter $\Lambda$ (MeV)} \\
               \cline{2-4}
                 &      414                     & 450           &    500            \\
\hline
\noalign{\smallskip}
HF contribution        & -35.507 & -32.786 & -25.066 \\
2nd order $pp$ diagram &  -5.736 &  -8.551 & -14.060 \\
3rd order $pp$ diagram &   0.017 &  -0.022 &   0.653 \\
3rd order $hh$ diagram &  -0.022 &  -0.021 &  -0.027 \\
3rd order $ph$ diagram &   1.040 &   1.200 &  -0.279 \\
\hline
\hline
\noalign{\smallskip}
\end{tabular*}
\end{table*}

\begin{table*}
\caption{Same as in Table \ref{tab2}, but including also 3NF effects.}
\label{tab3}
\smallskip
\begin{tabular*}{\textwidth}{@{\extracolsep{\fill}}cccc}
\hline
\hline
\noalign{\smallskip}
               & \multicolumn{3}{c}{Cutoff parameter $\Lambda$ (MeV)} \\
               \cline{2-4}
                 &      414                     & 450           &    500            \\
\hline
\noalign{\smallskip}
HF contribution        & -28.792 & -25.688 & -19.503 \\
2nd order $pp$ diagram &  -7.388 & -11.273 & -13.511 \\
3rd order $pp$ diagram &   0.563 &   0.745 &   1.642 \\
3rd order $hh$ diagram &  -0.010 &  -0.008 &  -0.008 \\
3rd order $ph$ diagram &   0.581 &   0.152 &  -1.516 \\
\hline
\hline
\noalign{\smallskip}
\end{tabular*}
\end{table*}

In spite of the fact that the regulator dependence is not removed, the
ability to obtain good saturation properties in a microscopic
calculation, where the parameters are determined {\it via} the
few-nucleon systems, should not be underestimated.

Another observation from this study is that the 3NF contribution to
the energy per nucleon in symmetric nuclear matter is larger than in
pure neutron matter~\cite{Coraggio13} (about a factor of two at
$k_F=1.35$fm$^{-1}$ for the $\Lambda=500$ MeV case). 
This may suggest that the weight of $3p-3h$ perturbative contributions induced 
by 3NF only (which are shown in Fig.~\ref{3p3h3NF} and are not
included here but come into play at second order and beyond) could be
non negligible.

\section{Concluding remarks and outlook}
\label{conc}
In this paper we have studied the regulator dependence of many-body
predictions of the EOS of symmetric nuclear matter, when employing
chiral two- and three-nucleon potentials.
This has been done within the framework of the perturbative Goldstone
expansion, and using three different cutoffs and regulator functions
for the derivation of the chiral potentials.
We have adopted a consistent choice of the LECs and of the regulator
functions for the two- and three-body components of the potential. 
In particular, the LECs $c_D$ and $c_E$ present in the 3NF have been
fixed as to reproduce the experimental $A=3$ binding energies and
Gamow-Teller matrix element in tritium $\beta$-decay.

Our calculations of the symmetric nuclear matter EOS show that,  when
employing chiral potentials with cutoffs $\Lambda=414$ and 450 MeV,
the regulator independence provided by the renormalization procedure
for the $A\leq3$ systems is preserved. 
We note again that these two potentials are found to exhibit good
perturbative behavior. 
Moreover, the introduction of 3NF effects proves to be crucial for
saturation and the predicted saturation properties are consistent 
with the empirical ones, within the uncertainty estimated on p.11.      
As mentioned above, this is a significant point, as it gives
confidence in an {\it ab initio} approach with two- and three-nucleon
forces consistent with each other {\it and} with the properties of
few-nucleon systems.

\begin{figure}[tb]
\begin{center}
\includegraphics[scale=1.0,angle=-90]{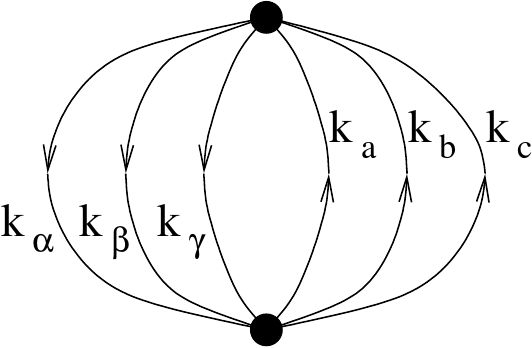}
\caption{Second-order Hugenholtz $3p-3h$ diagram of the Goldstone
  expansion with two 3NF vertices. Latin-letter subscripts denote
  particle states, Greek-letter subscripts correspond to hole states.}
\label{3p3h3NF}
\end{center}
\end{figure}

In a previous work \cite{Coraggio13}, where the same topic has been
studied for the pure neutron-matter EOS, we have found that the
inclusion of 3NF effects is crucial to restore the above regulator
independence also when employing the larger cutoff $\Lambda=500$ MeV
potential. 
This is not the case, at least within a perturbative approach, in
symmetric nuclear matter, where the $\Lambda=500$ MeV EOS is less
attractive than the other two by 3 MeV per nucleon around $k_F=1.35$
fm$^{-1}$.

From the observation made in the previous section about the relative
sizes of 3NF contributions in nuclear {\it vs.} neutron matter, we
conclude that a calculation of the second-order $3p-3h$ diagram may
shed light on whether the regulator dependence we have found is an
issue with the perturbative expansion or with higher-order terms in
ChPT, i.e. 3NF and 4NF at N$^3$LO
\cite{Epelbaum06,Ishikawa07,Bernard08,Bernard11}.

The inclusion of the diagram in Fig. \ref{3p3h3NF} will be a topic of
future studies, and may provide a better understanding of the
application of chiral interactions in microscopic nuclear structure
calculations.

\subsection*{Acknowledgments}
This work was supported in part by the U.S. Department of Energy under
Grant No.\ DE-FG02-03ER41270 and No.\ DE-FG02-97ER-41014. 
We thank Norbert Kaiser for helpful discussions concerning the
calculation of the third-order particle-hole diagram.

\bibliographystyle{apsrev}
\bibliography{biblio}

\begin{thebibliography}{48}
\expandafter\ifx\csname natexlab\endcsname\relax\def\natexlab#1{#1}\fi
\expandafter\ifx\csname bibnamefont\endcsname\relax
  \def\bibnamefont#1{#1}\fi
\expandafter\ifx\csname bibfnamefont\endcsname\relax
  \def\bibfnamefont#1{#1}\fi
\expandafter\ifx\csname citenamefont\endcsname\relax
  \def\citenamefont#1{#1}\fi
\expandafter\ifx\csname url\endcsname\relax
  \def\url#1{\texttt{#1}}\fi
\expandafter\ifx\csname urlprefix\endcsname\relax\def\urlprefix{URL }\fi
\providecommand{\bibinfo}[2]{#2}
\providecommand{\eprint}[2][]{\url{#2}}

\bibitem[{\citenamefont{Entem and Machleidt}(2003)}]{EM03}
\bibinfo{author}{\bibfnamefont{D.~R.} \bibnamefont{Entem}} \bibnamefont{and}
  \bibinfo{author}{\bibfnamefont{R.}~\bibnamefont{Machleidt}},
  \bibinfo{journal}{Phys. Rev. C} \textbf{\bibinfo{volume}{68}},
  \bibinfo{pages}{041001(R)} (\bibinfo{year}{2003}).

\bibitem[{\citenamefont{Epelbaum et~al.}(2005)\citenamefont{Epelbaum,
  Gl{\"o}ckle, and Meissner}}]{EGM05}
\bibinfo{author}{\bibfnamefont{E.}~\bibnamefont{Epelbaum}},
  \bibinfo{author}{\bibfnamefont{W.}~\bibnamefont{Gl{\"o}ckle}},
  \bibnamefont{and} \bibinfo{author}{\bibfnamefont{U.-G.}
  \bibnamefont{Meissner}}, \bibinfo{journal}{Nucl. Phys. A}
  \textbf{\bibinfo{volume}{747}}, \bibinfo{pages}{362} (\bibinfo{year}{2005}).

\bibitem[{\citenamefont{Machleidt and Entem}(2011)}]{ME11}
\bibinfo{author}{\bibfnamefont{R.}~\bibnamefont{Machleidt}} \bibnamefont{and}
  \bibinfo{author}{\bibfnamefont{D.~R.} \bibnamefont{Entem}},
  \bibinfo{journal}{Phys. Rep.} \textbf{\bibinfo{volume}{503}},
  \bibinfo{pages}{1} (\bibinfo{year}{2011}).

\bibitem[{\citenamefont{Weinberg}(1992)}]{Wei92}
\bibinfo{author}{\bibfnamefont{S.}~\bibnamefont{Weinberg}},
  \bibinfo{journal}{Phys. Lett. B} \textbf{\bibinfo{volume}{295}},
  \bibinfo{pages}{114} (\bibinfo{year}{1992}).

\bibitem[{\citenamefont{van Kolck}(1994)}]{Kol94}
\bibinfo{author}{\bibfnamefont{U.}~\bibnamefont{van Kolck}},
  \bibinfo{journal}{Phys. Rev. C} \textbf{\bibinfo{volume}{49}},
  \bibinfo{pages}{2932} (\bibinfo{year}{1994}).

\bibitem[{\citenamefont{Coraggio et~al.}(2002)\citenamefont{Coraggio, Covello,
  Gargano, N.Itaco, and Kuo}}]{Cor12}
\bibinfo{author}{\bibfnamefont{L.}~\bibnamefont{Coraggio}},
  \bibinfo{author}{\bibfnamefont{A.}~\bibnamefont{Covello}},
  \bibinfo{author}{\bibfnamefont{A.}~\bibnamefont{Gargano}},
  \bibinfo{author}{\bibnamefont{N.Itaco}}, \bibnamefont{and}
  \bibinfo{author}{\bibfnamefont{T.~T.~S.} \bibnamefont{Kuo}},
  \bibinfo{journal}{Ann. Phys.} \textbf{\bibinfo{volume}{327}},
  \bibinfo{pages}{2125} (\bibinfo{year}{2002}).

\bibitem[{\citenamefont{Coraggio et~al.}(2013)\citenamefont{Coraggio, Holt,
  Itaco, Machleidt, and Sammarruca}}]{Coraggio13}
\bibinfo{author}{\bibfnamefont{L.}~\bibnamefont{Coraggio}},
  \bibinfo{author}{\bibfnamefont{J.~W.} \bibnamefont{Holt}},
  \bibinfo{author}{\bibfnamefont{N.}~\bibnamefont{Itaco}},
  \bibinfo{author}{\bibfnamefont{R.}~\bibnamefont{Machleidt}},
  \bibnamefont{and}
  \bibinfo{author}{\bibfnamefont{F.}~\bibnamefont{Sammarruca}},
  \bibinfo{journal}{Phys. Rev. C} \textbf{\bibinfo{volume}{87}},
  \bibinfo{pages}{014322} (\bibinfo{year}{2013}).

\bibitem[{\citenamefont{Sammarruca et~al.}(2012)\citenamefont{Sammarruca, Chen,
  Coraggio, Itaco, and Machleidt}}]{Sammarruca12}
\bibinfo{author}{\bibfnamefont{F.}~\bibnamefont{Sammarruca}},
  \bibinfo{author}{\bibfnamefont{B.}~\bibnamefont{Chen}},
  \bibinfo{author}{\bibfnamefont{L.}~\bibnamefont{Coraggio}},
  \bibinfo{author}{\bibfnamefont{N.}~\bibnamefont{Itaco}}, \bibnamefont{and}
  \bibinfo{author}{\bibfnamefont{R.}~\bibnamefont{Machleidt}},
  \bibinfo{journal}{Phys. Rev. C} \textbf{\bibinfo{volume}{86}},
  \bibinfo{pages}{054317} (\bibinfo{year}{2012}).

\bibitem[{\citenamefont{Tews et~al.}(2013)\citenamefont{Tews, Kr\"uger,
  Hebeler, and Schwenk}}]{Tews13}
\bibinfo{author}{\bibfnamefont{I.}~\bibnamefont{Tews}},
  \bibinfo{author}{\bibfnamefont{T.}~\bibnamefont{Kr\"uger}},
  \bibinfo{author}{\bibfnamefont{K.}~\bibnamefont{Hebeler}}, \bibnamefont{and}
  \bibinfo{author}{\bibfnamefont{A.}~\bibnamefont{Schwenk}},
  \bibinfo{journal}{Phys. Rev. Lett.} \textbf{\bibinfo{volume}{110}},
  \bibinfo{pages}{032504} (\bibinfo{year}{2013}).

\bibitem[{\citenamefont{Holt et~al.}(2013)\citenamefont{Holt, Kaiser, and
  Weise}}]{HKW13}
\bibinfo{author}{\bibfnamefont{J.~W.} \bibnamefont{Holt}},
  \bibinfo{author}{\bibfnamefont{N.}~\bibnamefont{Kaiser}}, \bibnamefont{and}
  \bibinfo{author}{\bibfnamefont{W.}~\bibnamefont{Weise}},
  \bibinfo{journal}{Prog. Part. Nucl. Phys.} \textbf{\bibinfo{volume}{73}},
  \bibinfo{pages}{35} (\bibinfo{year}{2013}).

\bibitem[{\citenamefont{Kr\"uger et~al.}(2013)\citenamefont{Kr\"uger, Tews,
  Hebeler, and Schwenk}}]{Krueger13}
\bibinfo{author}{\bibfnamefont{T.}~\bibnamefont{Kr\"uger}},
  \bibinfo{author}{\bibfnamefont{I.}~\bibnamefont{Tews}},
  \bibinfo{author}{\bibfnamefont{K.}~\bibnamefont{Hebeler}}, \bibnamefont{and}
  \bibinfo{author}{\bibfnamefont{A.}~\bibnamefont{Schwenk}},
  \bibinfo{journal}{Phys. Rev. C} \textbf{\bibinfo{volume}{88}},
  \bibinfo{pages}{025802} (\bibinfo{year}{2013}).

\bibitem[{\citenamefont{Gezerlis et~al.}(2013)\citenamefont{Gezerlis, Tews,
  Epelbaum, Gandolfi, Hebeler, Nogga, and Schwenk}}]{Gezerlis13}
\bibinfo{author}{\bibfnamefont{A.}~\bibnamefont{Gezerlis}},
  \bibinfo{author}{\bibfnamefont{I.}~\bibnamefont{Tews}},
  \bibinfo{author}{\bibfnamefont{E.}~\bibnamefont{Epelbaum}},
  \bibinfo{author}{\bibfnamefont{S.}~\bibnamefont{Gandolfi}},
  \bibinfo{author}{\bibfnamefont{K.}~\bibnamefont{Hebeler}},
  \bibinfo{author}{\bibfnamefont{A.}~\bibnamefont{Nogga}}, \bibnamefont{and}
  \bibinfo{author}{\bibfnamefont{A.}~\bibnamefont{Schwenk}},
  \bibinfo{journal}{Phys. Rev. Lett.} \textbf{\bibinfo{volume}{111}},
  \bibinfo{pages}{032501} (\bibinfo{year}{2013}).

\bibitem[{\citenamefont{Carbone et~al.}(2013)\citenamefont{Carbone, Polls, and
  Rios}}]{Carbone13}
\bibinfo{author}{\bibfnamefont{A.}~\bibnamefont{Carbone}},
  \bibinfo{author}{\bibfnamefont{A.}~\bibnamefont{Polls}}, \bibnamefont{and}
  \bibinfo{author}{\bibfnamefont{A.}~\bibnamefont{Rios}},
  \bibinfo{journal}{Phys. Rev. C} \textbf{\bibinfo{volume}{88}},
  \bibinfo{pages}{044302} (\bibinfo{year}{2013}).

\bibitem[{\citenamefont{Baardsen et~al.}(2013)\citenamefont{Baardsen,
  Ekstr\"om, Hagen, and Hjorth-Jensen}}]{Baardsen13}
\bibinfo{author}{\bibfnamefont{G.}~\bibnamefont{Baardsen}},
  \bibinfo{author}{\bibfnamefont{A.}~\bibnamefont{Ekstr\"om}},
  \bibinfo{author}{\bibfnamefont{G.}~\bibnamefont{Hagen}}, \bibnamefont{and}
  \bibinfo{author}{\bibfnamefont{M.}~\bibnamefont{Hjorth-Jensen}},
  \bibinfo{journal}{Phys. Rev. C} \textbf{\bibinfo{volume}{88}},
  \bibinfo{pages}{054312} (\bibinfo{year}{2013}).

\bibitem[{\citenamefont{Kohno}(2013)}]{Kohno13}
\bibinfo{author}{\bibfnamefont{M.}~\bibnamefont{Kohno}},
  \bibinfo{journal}{Phys. Rev. C} \textbf{\bibinfo{volume}{88}},
  \bibinfo{pages}{064005} (\bibinfo{year}{2013}).

\bibitem[{\citenamefont{Hagen et~al.}(2014)\citenamefont{Hagen, Papenbrock,
  Ekstr\"om, Wendt, Baardsen, Gandolfi, Hjorth-Jensen, and Horowitz}}]{Hagen14}
\bibinfo{author}{\bibfnamefont{G.}~\bibnamefont{Hagen}},
  \bibinfo{author}{\bibfnamefont{T.}~\bibnamefont{Papenbrock}},
  \bibinfo{author}{\bibfnamefont{A.}~\bibnamefont{Ekstr\"om}},
  \bibinfo{author}{\bibfnamefont{K.~A.} \bibnamefont{Wendt}},
  \bibinfo{author}{\bibfnamefont{G.}~\bibnamefont{Baardsen}},
  \bibinfo{author}{\bibfnamefont{S.}~\bibnamefont{Gandolfi}},
  \bibinfo{author}{\bibfnamefont{M.}~\bibnamefont{Hjorth-Jensen}},
  \bibnamefont{and} \bibinfo{author}{\bibfnamefont{C.~J.}
  \bibnamefont{Horowitz}}, \bibinfo{journal}{Phys. Rev. C}
  \textbf{\bibinfo{volume}{89}}, \bibinfo{pages}{014319}
  (\bibinfo{year}{2014}).

\bibitem[{\citenamefont{Hebeler and Schwenk}(2010)}]{HS10}
\bibinfo{author}{\bibfnamefont{K.}~\bibnamefont{Hebeler}} \bibnamefont{and}
  \bibinfo{author}{\bibfnamefont{A.}~\bibnamefont{Schwenk}},
  \bibinfo{journal}{Phys. Rev. C} \textbf{\bibinfo{volume}{82}},
  \bibinfo{pages}{014314} (\bibinfo{year}{2010}).

\bibitem[{\citenamefont{G\aa{}rdestig and Phillips}(2006)}]{Gardestig06}
\bibinfo{author}{\bibfnamefont{A.}~\bibnamefont{G\aa{}rdestig}}
  \bibnamefont{and} \bibinfo{author}{\bibfnamefont{D.~R.}
  \bibnamefont{Phillips}}, \bibinfo{journal}{Phys. Rev. Lett.}
  \textbf{\bibinfo{volume}{96}}, \bibinfo{pages}{232301}
  (\bibinfo{year}{2006}).

\bibitem[{\citenamefont{Gazit}(2008)}]{Gazit08}
\bibinfo{author}{\bibfnamefont{D.}~\bibnamefont{Gazit}},
  \bibinfo{journal}{Phys. Lett. B} \textbf{\bibinfo{volume}{666}},
  \bibinfo{pages}{472} (\bibinfo{year}{2008}).

\bibitem[{\citenamefont{Marcucci et~al.}(2012)\citenamefont{Marcucci, Kievsky,
  Rosati, Schiavilla, and Viviani}}]{Marcucci12}
\bibinfo{author}{\bibfnamefont{L.~E.} \bibnamefont{Marcucci}},
  \bibinfo{author}{\bibfnamefont{A.}~\bibnamefont{Kievsky}},
  \bibinfo{author}{\bibfnamefont{S.}~\bibnamefont{Rosati}},
  \bibinfo{author}{\bibfnamefont{R.}~\bibnamefont{Schiavilla}},
  \bibnamefont{and} \bibinfo{author}{\bibfnamefont{M.}~\bibnamefont{Viviani}},
  \bibinfo{journal}{Phys. Rev. Lett.} \textbf{\bibinfo{volume}{108}},
  \bibinfo{pages}{052502} (\bibinfo{year}{2012}).

\bibitem[{\citenamefont{Coraggio et~al.}(2007)\citenamefont{Coraggio, Covello,
  Gargano, Itaco, Kuo, Entem, and Machleidt}}]{Cor07}
\bibinfo{author}{\bibfnamefont{L.}~\bibnamefont{Coraggio}},
  \bibinfo{author}{\bibfnamefont{A.}~\bibnamefont{Covello}},
  \bibinfo{author}{\bibfnamefont{A.}~\bibnamefont{Gargano}},
  \bibinfo{author}{\bibfnamefont{N.}~\bibnamefont{Itaco}},
  \bibinfo{author}{\bibfnamefont{T.~T.~S.} \bibnamefont{Kuo}},
  \bibinfo{author}{\bibfnamefont{D.~R.} \bibnamefont{Entem}}, \bibnamefont{and}
  \bibinfo{author}{\bibfnamefont{R.}~\bibnamefont{Machleidt}},
  \bibinfo{journal}{Phys. Rev. C} \textbf{\bibinfo{volume}{75}},
  \bibinfo{pages}{024311} (\bibinfo{year}{2007}).

\bibitem[{\citenamefont{Viviani et~al.}(2013)\citenamefont{Viviani, Girlanda,
  Kievsky, and Marcucci}}]{Viv13}
\bibinfo{author}{\bibfnamefont{M.}~\bibnamefont{Viviani}},
  \bibinfo{author}{\bibfnamefont{L.}~\bibnamefont{Girlanda}},
  \bibinfo{author}{\bibfnamefont{A.}~\bibnamefont{Kievsky}}, \bibnamefont{and}
  \bibinfo{author}{\bibfnamefont{L.~E.} \bibnamefont{Marcucci}},
  \bibinfo{journal}{Phys. Rev. Lett.} \textbf{\bibinfo{volume}{111}},
  \bibinfo{pages}{172302} (\bibinfo{year}{2013}).

\bibitem[{\citenamefont{Piarulli et~al.}(2013)\citenamefont{Piarulli, Girlanda,
  Marcucci, Pastore, Schiavilla, and Viviani}}]{Pia13}
\bibinfo{author}{\bibfnamefont{M.}~\bibnamefont{Piarulli}},
  \bibinfo{author}{\bibfnamefont{L.}~\bibnamefont{Girlanda}},
  \bibinfo{author}{\bibfnamefont{L.~E.} \bibnamefont{Marcucci}},
  \bibinfo{author}{\bibfnamefont{S.}~\bibnamefont{Pastore}},
  \bibinfo{author}{\bibfnamefont{R.}~\bibnamefont{Schiavilla}},
  \bibnamefont{and} \bibinfo{author}{\bibfnamefont{M.}~\bibnamefont{Viviani}},
  \bibinfo{journal}{Phys. Rev. C} \textbf{\bibinfo{volume}{87}},
  \bibinfo{pages}{014006} (\bibinfo{year}{2013}).

\bibitem[{\citenamefont{Marcucci et~al.}(2013)\citenamefont{Marcucci,
  Schiavilla, and Viviani}}]{Mar13}
\bibinfo{author}{\bibfnamefont{L.~E.} \bibnamefont{Marcucci}},
  \bibinfo{author}{\bibfnamefont{R.}~\bibnamefont{Schiavilla}},
  \bibnamefont{and} \bibinfo{author}{\bibfnamefont{M.}~\bibnamefont{Viviani}},
  \bibinfo{journal}{Phys. Rev. Lett.} \textbf{\bibinfo{volume}{110}},
  \bibinfo{pages}{192503} (\bibinfo{year}{2013}).

\bibitem[{\citenamefont{Hebeler et~al.}(2011)\citenamefont{Hebeler, Bogner,
  Furnstahl, Nogga, and Schwenk}}]{Hebeler11}
\bibinfo{author}{\bibfnamefont{K.}~\bibnamefont{Hebeler}},
  \bibinfo{author}{\bibfnamefont{S.~K.} \bibnamefont{Bogner}},
  \bibinfo{author}{\bibfnamefont{R.~J.} \bibnamefont{Furnstahl}},
  \bibinfo{author}{\bibfnamefont{A.}~\bibnamefont{Nogga}}, \bibnamefont{and}
  \bibinfo{author}{\bibfnamefont{A.}~\bibnamefont{Schwenk}},
  \bibinfo{journal}{Phys. Rev. C} \textbf{\bibinfo{volume}{83}},
  \bibinfo{pages}{031301(R)} (\bibinfo{year}{2011}).

\bibitem[{\citenamefont{Holt et~al.}(2009)\citenamefont{Holt, Kaiser, and
  Weise}}]{HKW09}
\bibinfo{author}{\bibfnamefont{J.~W.} \bibnamefont{Holt}},
  \bibinfo{author}{\bibfnamefont{N.}~\bibnamefont{Kaiser}}, \bibnamefont{and}
  \bibinfo{author}{\bibfnamefont{W.}~\bibnamefont{Weise}},
  \bibinfo{journal}{Phys. Rev. C} \textbf{\bibinfo{volume}{79}},
  \bibinfo{pages}{054331} (\bibinfo{year}{2009}).

\bibitem[{\citenamefont{Holt et~al.}(2010)\citenamefont{Holt, Kaiser, and
  Weise}}]{HKW10}
\bibinfo{author}{\bibfnamefont{J.~W.} \bibnamefont{Holt}},
  \bibinfo{author}{\bibfnamefont{N.}~\bibnamefont{Kaiser}}, \bibnamefont{and}
  \bibinfo{author}{\bibfnamefont{W.}~\bibnamefont{Weise}},
  \bibinfo{journal}{Phys. Rev. C} \textbf{\bibinfo{volume}{81}},
  \bibinfo{pages}{024002} (\bibinfo{year}{2010}).

\bibitem[{\citenamefont{Epelbaum et~al.}(2009)\citenamefont{Epelbaum, Hammer,
  and Mei\ss{}ner}}]{EHM09}
\bibinfo{author}{\bibfnamefont{E.}~\bibnamefont{Epelbaum}},
  \bibinfo{author}{\bibfnamefont{H.-W.} \bibnamefont{Hammer}},
  \bibnamefont{and} \bibinfo{author}{\bibfnamefont{U.-G.}
  \bibnamefont{Mei\ss{}ner}}, \bibinfo{journal}{Rev. Mod. Phys.}
  \textbf{\bibinfo{volume}{81}}, \bibinfo{pages}{1773} (\bibinfo{year}{2009}).

\bibitem[{\citenamefont{Stoks et~al.}(1993)\citenamefont{Stoks, Klomp,
  Rentmeester, and de~Swart}}]{Sto93}
\bibinfo{author}{\bibfnamefont{V.~G.~J.} \bibnamefont{Stoks}},
  \bibinfo{author}{\bibfnamefont{R.~A.~M.} \bibnamefont{Klomp}},
  \bibinfo{author}{\bibfnamefont{M.~C.~M.} \bibnamefont{Rentmeester}},
  \bibnamefont{and} \bibinfo{author}{\bibfnamefont{J.~J.}
  \bibnamefont{de~Swart}}, \bibinfo{journal}{Phys. Rev. C}
  \textbf{\bibinfo{volume}{48}}, \bibinfo{pages}{792} (\bibinfo{year}{1993}).

\bibitem[{\citenamefont{Arndt et~al.}()\citenamefont{Arndt, Strakovsky, and
  Workman}}]{SM99}
\bibinfo{author}{\bibfnamefont{R.~A.} \bibnamefont{Arndt}},
  \bibinfo{author}{\bibfnamefont{I.~I.} \bibnamefont{Strakovsky}},
  \bibnamefont{and} \bibinfo{author}{\bibfnamefont{R.~L.}
  \bibnamefont{Workman}}, \emph{\bibinfo{title}{Said scattering analysis
  interactive dial-in computer facility, george washington university (formerly
  virginia polytechnic institute), solution sm99 (summer 1999)}}.

\bibitem[{\citenamefont{Weinberg}(1979)}]{Wei79}
\bibinfo{author}{\bibfnamefont{S.}~\bibnamefont{Weinberg}},
  \bibinfo{journal}{Physica} \textbf{\bibinfo{volume}{96A}},
  \bibinfo{pages}{327} (\bibinfo{year}{1979}).

\bibitem[{\citenamefont{Navr\'atil et~al.}(2007)\citenamefont{Navr\'atil,
  Gueorguiev, Vary, Ormand, and Nogga}}]{Nav07}
\bibinfo{author}{\bibfnamefont{P.}~\bibnamefont{Navr\'atil}},
  \bibinfo{author}{\bibfnamefont{V.~G.} \bibnamefont{Gueorguiev}},
  \bibinfo{author}{\bibfnamefont{J.~P.} \bibnamefont{Vary}},
  \bibinfo{author}{\bibfnamefont{W.~E.} \bibnamefont{Ormand}},
  \bibnamefont{and} \bibinfo{author}{\bibfnamefont{A.}~\bibnamefont{Nogga}},
  \bibinfo{journal}{Phys. Rev. Lett.} \textbf{\bibinfo{volume}{99}},
  \bibinfo{pages}{042501} (\bibinfo{year}{2007}).

\bibitem[{\citenamefont{Marcucci et~al.}(2000)\citenamefont{Marcucci,
  Schiavilla, Viviani, Kievsky, Rosati, and Beacom}}]{Mar00}
\bibinfo{author}{\bibfnamefont{L.~E.} \bibnamefont{Marcucci}},
  \bibinfo{author}{\bibfnamefont{R.}~\bibnamefont{Schiavilla}},
  \bibinfo{author}{\bibfnamefont{M.}~\bibnamefont{Viviani}},
  \bibinfo{author}{\bibfnamefont{A.}~\bibnamefont{Kievsky}},
  \bibinfo{author}{\bibfnamefont{S.}~\bibnamefont{Rosati}}, \bibnamefont{and}
  \bibinfo{author}{\bibfnamefont{J.~F.} \bibnamefont{Beacom}},
  \bibinfo{journal}{Phys. Rev. C} \textbf{\bibinfo{volume}{63}},
  \bibinfo{pages}{015801} (\bibinfo{year}{2000}).

\bibitem[{\citenamefont{Park et~al.}(2003)\citenamefont{Park, Marcucci,
  Schiavilla, Viviani, Kievsky, Rosati, Kubodera, Min, and Rho}}]{Par03}
\bibinfo{author}{\bibfnamefont{T.-S.} \bibnamefont{Park}},
  \bibinfo{author}{\bibfnamefont{L.~E.} \bibnamefont{Marcucci}},
  \bibinfo{author}{\bibfnamefont{R.}~\bibnamefont{Schiavilla}},
  \bibinfo{author}{\bibfnamefont{M.}~\bibnamefont{Viviani}},
  \bibinfo{author}{\bibfnamefont{A.}~\bibnamefont{Kievsky}},
  \bibinfo{author}{\bibfnamefont{S.}~\bibnamefont{Rosati}},
  \bibinfo{author}{\bibfnamefont{K.}~\bibnamefont{Kubodera}},
  \bibinfo{author}{\bibfnamefont{D.-P.} \bibnamefont{Min}}, \bibnamefont{and}
  \bibinfo{author}{\bibfnamefont{M.}~\bibnamefont{Rho}},
  \bibinfo{journal}{Phys. Rev. C} \textbf{\bibinfo{volume}{67}},
  \bibinfo{pages}{055206} (\bibinfo{year}{2003}).

\bibitem[{\citenamefont{Marcucci et~al.}(2011)\citenamefont{Marcucci, Piarulli,
  Viviani, Girlanda, Kievsky, Rosati, and Schiavilla}}]{Mar11}
\bibinfo{author}{\bibfnamefont{L.~E.} \bibnamefont{Marcucci}},
  \bibinfo{author}{\bibfnamefont{M.}~\bibnamefont{Piarulli}},
  \bibinfo{author}{\bibfnamefont{M.}~\bibnamefont{Viviani}},
  \bibinfo{author}{\bibfnamefont{L.}~\bibnamefont{Girlanda}},
  \bibinfo{author}{\bibfnamefont{A.}~\bibnamefont{Kievsky}},
  \bibinfo{author}{\bibfnamefont{S.}~\bibnamefont{Rosati}}, \bibnamefont{and}
  \bibinfo{author}{\bibfnamefont{R.}~\bibnamefont{Schiavilla}},
  \bibinfo{journal}{Phys. Rev. C} \textbf{\bibinfo{volume}{83}},
  \bibinfo{pages}{014002} (\bibinfo{year}{2011}).

\bibitem[{\citenamefont{Kievsky et~al.}(2008)\citenamefont{Kievsky, Rosati,
  Viviani, Marcucci, and Girlanda}}]{Kie08}
\bibinfo{author}{\bibfnamefont{A.}~\bibnamefont{Kievsky}},
  \bibinfo{author}{\bibfnamefont{S.}~\bibnamefont{Rosati}},
  \bibinfo{author}{\bibfnamefont{M.}~\bibnamefont{Viviani}},
  \bibinfo{author}{\bibfnamefont{L.~E.} \bibnamefont{Marcucci}},
  \bibnamefont{and} \bibinfo{author}{\bibfnamefont{L.}~\bibnamefont{Girlanda}},
  \bibinfo{journal}{Journal of Physics G: Nuclear and Particle Physics}
  \textbf{\bibinfo{volume}{35}}, \bibinfo{pages}{063101}
  (\bibinfo{year}{2008}).

\bibitem[{\citenamefont{Bernard
  et~al.}(2008{\natexlab{a}})\citenamefont{Bernard, Epelbaum, Krebs, and
  Meissner}}]{Ber08}
\bibinfo{author}{\bibfnamefont{V.}~\bibnamefont{Bernard}},
  \bibinfo{author}{\bibfnamefont{E.}~\bibnamefont{Epelbaum}},
  \bibinfo{author}{\bibfnamefont{H.}~\bibnamefont{Krebs}}, \bibnamefont{and}
  \bibinfo{author}{\bibfnamefont{U.-G.} \bibnamefont{Meissner}},
  \bibinfo{journal}{Phys.Rev.} \textbf{\bibinfo{volume}{C77}},
  \bibinfo{pages}{064004} (\bibinfo{year}{2008}{\natexlab{a}}),
  \eprint{0712.1967}.

\bibitem[{\citenamefont{Bernard
  et~al.}(2011{\natexlab{a}})\citenamefont{Bernard, Epelbaum, Krebs, and
  Meissner}}]{Ber11}
\bibinfo{author}{\bibfnamefont{V.}~\bibnamefont{Bernard}},
  \bibinfo{author}{\bibfnamefont{E.}~\bibnamefont{Epelbaum}},
  \bibinfo{author}{\bibfnamefont{H.}~\bibnamefont{Krebs}}, \bibnamefont{and}
  \bibinfo{author}{\bibfnamefont{U.-G.} \bibnamefont{Meissner}},
  \bibinfo{journal}{Phys.Rev.} \textbf{\bibinfo{volume}{C84}},
  \bibinfo{pages}{054001} (\bibinfo{year}{2011}{\natexlab{a}}),
  \eprint{1108.3816}.

\bibitem[{\citenamefont{Skibinski et~al.}(2013)\citenamefont{Skibinski, Golak,
  Topolnicki, Witala, Epelbaum et~al.}}]{Ski13}
\bibinfo{author}{\bibfnamefont{R.}~\bibnamefont{Skibinski}},
  \bibinfo{author}{\bibfnamefont{J.}~\bibnamefont{Golak}},
  \bibinfo{author}{\bibfnamefont{K.}~\bibnamefont{Topolnicki}},
  \bibinfo{author}{\bibfnamefont{H.}~\bibnamefont{Witala}},
  \bibinfo{author}{\bibfnamefont{E.}~\bibnamefont{Epelbaum}},
  \bibnamefont{et~al.}, \bibinfo{journal}{Few Body Syst.}
  \textbf{\bibinfo{volume}{54}}, \bibinfo{pages}{1315} (\bibinfo{year}{2013}).

\bibitem[{\citenamefont{Witala et~al.}(2013)\citenamefont{Witala, Golak,
  Skibinski, Topolnicki, Kamada et~al.}}]{Wit13}
\bibinfo{author}{\bibfnamefont{H.}~\bibnamefont{Witala}},
  \bibinfo{author}{\bibfnamefont{J.}~\bibnamefont{Golak}},
  \bibinfo{author}{\bibfnamefont{R.}~\bibnamefont{Skibinski}},
  \bibinfo{author}{\bibfnamefont{K.}~\bibnamefont{Topolnicki}},
  \bibinfo{author}{\bibfnamefont{H.}~\bibnamefont{Kamada}},
  \bibnamefont{et~al.}, \bibinfo{journal}{Few Body Syst.}
  \textbf{\bibinfo{volume}{54}}, \bibinfo{pages}{897} (\bibinfo{year}{2013}).

\bibitem[{\citenamefont{Bethe}(1965)}]{Bethe65}
\bibinfo{author}{\bibfnamefont{H.~A.} \bibnamefont{Bethe}},
  \bibinfo{journal}{Phys. Rev.} \textbf{\bibinfo{volume}{138}},
  \bibinfo{pages}{B804} (\bibinfo{year}{1965}).

\bibitem[{\citenamefont{MacKenzie}(1969)}]{MacKenzie69}
\bibinfo{author}{\bibfnamefont{J.~J.} \bibnamefont{MacKenzie}},
  \bibinfo{journal}{Phys. Rev.} \textbf{\bibinfo{volume}{179}},
  \bibinfo{pages}{1002} (\bibinfo{year}{1969}).

\bibitem[{\citenamefont{Ekstr\"om et~al.}(2013)\citenamefont{Ekstr\"om,
  Baardsen, Forss\'en, Hagen, Hjorth-Jensen, Jansen, Machleidt, Nazarewicz,
  Papenbrock, Sarich et~al.}}]{Ekstrom13}
\bibinfo{author}{\bibfnamefont{A.}~\bibnamefont{Ekstr\"om}},
  \bibinfo{author}{\bibfnamefont{G.}~\bibnamefont{Baardsen}},
  \bibinfo{author}{\bibfnamefont{C.}~\bibnamefont{Forss\'en}},
  \bibinfo{author}{\bibfnamefont{G.}~\bibnamefont{Hagen}},
  \bibinfo{author}{\bibfnamefont{M.}~\bibnamefont{Hjorth-Jensen}},
  \bibinfo{author}{\bibfnamefont{G.~R.} \bibnamefont{Jansen}},
  \bibinfo{author}{\bibfnamefont{R.}~\bibnamefont{Machleidt}},
  \bibinfo{author}{\bibfnamefont{W.}~\bibnamefont{Nazarewicz}},
  \bibinfo{author}{\bibfnamefont{T.}~\bibnamefont{Papenbrock}},
  \bibinfo{author}{\bibfnamefont{J.}~\bibnamefont{Sarich}},
  \bibnamefont{et~al.}, \bibinfo{journal}{Phys. Rev. Lett.}
  \textbf{\bibinfo{volume}{110}}, \bibinfo{pages}{192502}
  (\bibinfo{year}{2013}).

\bibitem[{\citenamefont{Baker and Gammel}(1970)}]{BG70}
\bibinfo{author}{\bibfnamefont{G.~A.} \bibnamefont{Baker}} \bibnamefont{and}
  \bibinfo{author}{\bibfnamefont{J.~L.} \bibnamefont{Gammel}},
  \emph{\bibinfo{title}{The Pad{\'e} Approximant in Theoretical Physics}},
  vol.~\bibinfo{volume}{71} of \emph{\bibinfo{series}{Mathematics in Science
  and Engineering}} (\bibinfo{publisher}{Academic Press, New York},
  \bibinfo{year}{1970}).

\bibitem[{\citenamefont{Epelbaum}(2006)}]{Epelbaum06}
\bibinfo{author}{\bibfnamefont{E.}~\bibnamefont{Epelbaum}},
  \bibinfo{journal}{Phys. Lett. B} \textbf{\bibinfo{volume}{639}},
  \bibinfo{pages}{456} (\bibinfo{year}{2006}).

\bibitem[{\citenamefont{Ishikawa and Robilotta}(2007)}]{Ishikawa07}
\bibinfo{author}{\bibfnamefont{S.}~\bibnamefont{Ishikawa}} \bibnamefont{and}
  \bibinfo{author}{\bibfnamefont{M.~R.} \bibnamefont{Robilotta}},
  \bibinfo{journal}{Phys. Rev. C} \textbf{\bibinfo{volume}{76}},
  \bibinfo{pages}{014006} (\bibinfo{year}{2007}).

\bibitem[{\citenamefont{Bernard
  et~al.}(2008{\natexlab{b}})\citenamefont{Bernard, Epelbaum, Krebs, and
  Mei\ss{}ner}}]{Bernard08}
\bibinfo{author}{\bibfnamefont{V.}~\bibnamefont{Bernard}},
  \bibinfo{author}{\bibfnamefont{E.}~\bibnamefont{Epelbaum}},
  \bibinfo{author}{\bibfnamefont{H.}~\bibnamefont{Krebs}}, \bibnamefont{and}
  \bibinfo{author}{\bibfnamefont{U.-G.} \bibnamefont{Mei\ss{}ner}},
  \bibinfo{journal}{Phys. Rev. C} \textbf{\bibinfo{volume}{77}},
  \bibinfo{pages}{064004} (\bibinfo{year}{2008}{\natexlab{b}}).

\bibitem[{\citenamefont{Bernard
  et~al.}(2011{\natexlab{b}})\citenamefont{Bernard, Epelbaum, Krebs, and
  Mei\ss{}ner}}]{Bernard11}
\bibinfo{author}{\bibfnamefont{V.}~\bibnamefont{Bernard}},
  \bibinfo{author}{\bibfnamefont{E.}~\bibnamefont{Epelbaum}},
  \bibinfo{author}{\bibfnamefont{H.}~\bibnamefont{Krebs}}, \bibnamefont{and}
  \bibinfo{author}{\bibfnamefont{U.-G.} \bibnamefont{Mei\ss{}ner}},
  \bibinfo{journal}{Phys. Rev. C} \textbf{\bibinfo{volume}{84}},
  \bibinfo{pages}{054001} (\bibinfo{year}{2011}{\natexlab{b}}).

\end{thebibliography}

\end{document}